# Ionic Modulation of Interfacial Magnetism through Electrostatic Doping in Pt/YIG bilayer heterostructure

*Mengmeng Guan[#], Lei Wang[#], Ziyao Zhou[*], Guohua Dong, Shishun Zhao, Wei Su, Tai Min, Jing Ma, Zhongqiang Hu, Wei Ren, Zuo-Guang Ye, Ce-Wen Nan, Ming Liu[*]*

Mengmeng Guan, Prof. Ziyao Zhou, Guohua Dong, Shishun Zhao, Wei Su, Prof. Zhongqiang Hu, Prof. Wei Ren, Prof. Zuo-Guang Ye, Prof. Ming Liu

Electronic Materials Research Laboratory, Key Laboratory of the Ministry of Education and International Center for Dielectric Research School of Electronic and Information Engineering, Xi'an Jiaotong University, Xi'an, Shaanxi, 710049, China

Dr. Lei Wang, Prof. Tai Min

Center for Spintronics and Quantum System, State Key Laboratory for Mechanical Behavior of Materials, School of Materials Science and Engineering, Xi'an Jiaotong University, Xi'an, Shaanxi, 710049, China

Prof. Jing Ma, Prof Ce-Wen Nan

State Key Lab of New Ceramics and Fine Processing, School of Materials Science and Engineering, Tsinghua University, Beijing, 100084, China.

Prof. Zuo-Guang Ye

Department of Chemistry and 4D LABS, Simon Fraser University, Burnaby, British Columbia, V5A 1S6, Canada

[#]These authors contributed equally

[*]E-mail: ziyaozhou@xjtu.edu.cn; mingliu@xjtu.edu.cn

Keywords: Yttrium iron garnet, ionic liquid gating, ferromagnetic resonance, magnetic proximity effect, magnetoelectric coupling

**Abstract**

Voltage modulation of yttrium iron garnet (YIG) with compactness, high speed response, energy efficiency and both practical/theoretical siginificances can be widely applied to various YIG based spintronics such as spin Hall, spin pumping, spin Seeback effects. Here we initial an ionic modulation of interfacial magnetism process on YIG/Pt bilayer heterostructures, where the Pt capping would influence the ferromagnetic (FMR) field position significantly, and realize a significant magnetism enhancement in bilayer system. A large voltage induced FMR field shifts of 690 Oe has been achieved in YIG (13 nm)/Pt (3 nm) multilayer heterostructures under a small voltage bias of 4.5 V. The remarkable ME tunability comes from voltage induced extra FM ordering in Pt metal layer near the Pt/YIG interface. The first-principle theoretical simulation reveal that the electrostatic doping induced $Pt^{5+}$ ions have strong magnetic ordering due to uncompensated d orbit electrons. The large voltage control of FMR change pave a foundation towards novel voltage tunable YIG based spintronics.

**Introduction**

Yttrium iron garnet ($Y_3Fe_5O_{12}$, YIG), a commonly used magnetic material, has high Curie temperature ($T_C \sim 650$ K), very low intrinsic damping ($\alpha \sim 10^{-5}$), long spin transmitting length (~1 cm), broad band gap ($E_g \sim 2.85$ eV), and very low ferromagnetic resonance (FMR) linewidth (~1 Oe).[1-3] It is an ideal ferromagnetic insolator, which plays a key role in spintronics devices and exhibits a plentiful of spintronic behaviors such as spin pumping[4], spin Hall[5], spin Seebeck[4,6], and magnetic proximity effects (MPE)[7,8] etc. Recently, the research interests in spin-orbital torque (SOT) were focused on interface between heavy metal and magnetic metals/insulators, in particular, current-driven SOT in YIG/heavy metal heterostructures.[9-16] In 2013, Sun et al. discovered that Pt thin film (>3 nm) capping onto YIG layer led to a damping change and an accompanied strong FMR shift due to magnetic proximity effect (MPE), where the ferromagnetic (FM) ordering in the Pt layer near the YIG/Pt interface was created by dynamic exchange coupling.[16] Nevertheless, the most recent progresses in this field focused on studying spin behaviors in non-magnetic layer in heavy metal (HM)/YIG bilayer heterostructures, where YIG serves as spin generator; while few researches discussed how the interfacial effect influences YIG thin film property as well as the total magnetism in the whole bilayer system.

Moreover, if the interfacial effect between Pt and YIG, for example, can be modified by a localized electric field (E-field) because of possible voltage induced Fermi level shift, we can therefore overcome the state of the art challenges of voltage modulation of spin phenomena in a fast, compact and energy efficient way, instead of current, in YIG related heterostructures. This voltage modulation approach also provides an extra E-manipulation degree of freedom for spintronics community, for instance, voltage controllable spin Hall, spin pumping, SOT effects. Here we propose an ionic liquid (IL) gating approach for ferromagnetism modulation, where IL serves as an effective gating media that provides significant interfacial charge accumulation under E-field.[17-21] IL gating manipulates the interfacial magnetism of ultrathin metallic films by changing the electron density at the Fermi level,[18] thereby modulates magnetism of oxide thin films through changing oxygen vacancies,[19, 20] and even triggers the tri-phase transformation in some oxides by controlled ionic doping.[21] It has many benefits over conventional multiferroics such as room temperature operation, low gating voltage ($V_g$) (<5 V), high ME tunability[22] and compatibility among various substrates.

In this work, a series of YIG thin films with different thicknesses (7 to 35 nm) were epitaxially deposited onto GGG single crystal substrates. Heavy-metal Pt thin films were then coated onto them to serve as one of the electrode of the gating process and the coupling media between the YIG layer and IL. By placing IL onto these YIG/Pt heterostructures, we established an ideal Field Effect Transistor (FET) structure and then applied a small $V_g$ (<5 V) across the IL layer. A large voltage-induced FMR field shift of 690 Oe was obtained in 13 nm YIG film at -110 ℃. The first principle calculation demonstrates that the enhancement of spin ordering and corresponding FMR field shift are resulted from electrical induced extra FM ordering in Pt metal layer during the IL gating process[172]. The first principle calculation shows that the uncompensated d orbital electrons of $Pt^{5+}$ and shifted Fermi level under E-bias is the origin of the extra FM ordering. We believe that this relative new ME gating mechanism – the ionic created FM ordering in heavy metal layer will attract further research attentions because Pt/YIG (or other heavy metal/YIG) as well-known SOT systems is still a research hotspot nowadays. Additionally, YIG is a perfect media in radio frequency and microwave devices such as filters,[23] shifters,[24] isolators,[25] circulators, spin wave components[9,10,26] and ultra-lower-power dissipation devices,[27] as well as in optical devices.[28] Electric field (E-field) manipulation of YIG thereby is of great significance to obtain voltage-tunable YIG devices with compactness, high-speed response, energy efficiency and extra degrees of manipulation freedom,[23,29,30] The large FMR tunability in YIG based heterostructures also appears a great application potential in tunable RF/microwave devices like bandpass filters and tunable spintronics/magnonics devices such as spin wave transisters.

**Results**

**Magnetic proximity effect in YIG/Pt.** The YIG layers were deposited on (111) GGG substrates using pulsed laser deposition (PLD) method with various thicknesses of 7 nm, 13 nm and 35 nm. The YIG film (35 nm) was chosen for the structure analysis for it yielded a stronger signal. Figure 1(a) shows the XRD pattern of the 35 nm YIG sample, which indicates that the YIG film was (111)-orientated. Figure 1(b) is the cross-section TEM image of the GGG/YIG (35 nm)/Pt (3 nm) sample, showing a good epitaxy of the YIG thin film. The in-plane magnetic hysteresis of the YIG (35 nm) sample before and after Pt (3 nm) capping is summarized in Figure 1(c). The hysteresis loop after Pt capping becomes more quadrate, indicating a strong coupling (MPE) between the YIG and Pt layers. The coupling effect induces an effective FM ordering and enhances the magnetization accordingly. Electron spin resonance (ESR) method is a very powerful tool to quantitatively determine the spatial magnetic anisotropy of these samples. To explore the origin of the magnetic coupling in these heterostructures, we also investigated the YIG thickness dependence of the FMR field ($H_r$) shift by the ESR technicque. Figure 1(d) shows the schematic of the sample in the ESR microwave cavity, the sample can be rotated with the sample hold to varify the magnetic field direction, the magnetic moment in the sample resonace when the magnetic field H equals $H_r$, which can be defined with the Kittel formula $f = (\gamma/2\pi)\sqrt{H_r^2 + H_r M_s}$ for the in plane condition and $f = (\gamma/2\pi)(H_r - M_s)$ for the out of plane condition.[31] where $f$ is the frequency of the microwave in the cavity, $\gamma/2\pi \triangleq 28$ (GHz/10 kOe) is the literature value of the gyromagnetic ratio[32], $M_s$ is the saturation magnetization. We can also dedicate from the Kittel formula that the $H_r$ will shift when $M_s$ changed. Figure 1(e) shows FMR spectra of bare YIG (35 nm) and YIG (35 nm) capped with Pt (3 nm). With Pt capping, we notice that the in-plane $H_r$ becomes smaller and the out-of-plane $H_r$ becomes larger, giving rise to an enhanced FM ordering and equavalent increased in-plane magnetic anisotropy, which correspond with the hysteresis loop change after Pt capping. As shown in Figure 1(f), the $H_r$ shifts along the in-plane and out-of-plane directions represent a YIG layer thickness dependence. We attribute this interfacial phenomenon to the MPE that comes from the interfacial coupling of the YIG layer and Pt layer and creates of FM ordering at the interface accordingly.

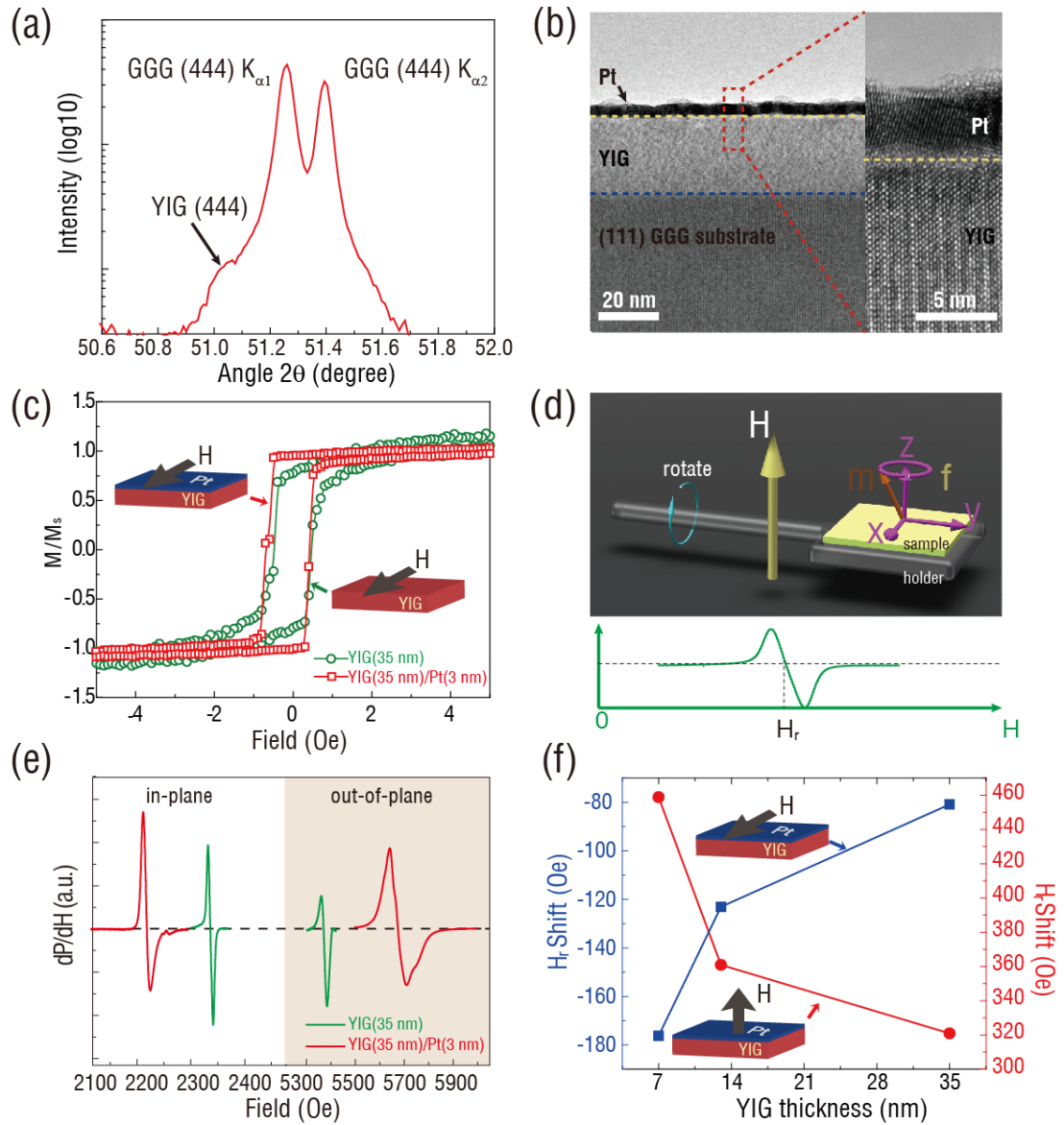

**Figure 1. Magnetic proximity effect in YIG/Pt.** (a) X-Ray diffraction of the GGG/YIG (35 nm)/Pt (3 nm) sample, (b) the cross section TEM of the same sample. (c) In-plane normalized magnetic hysteresis loops of the YIG (35 nm)/Pt (3 nm) (red) and bare YIG (35 nm) (green) samples. (d) schematic of the sample in the ESR cavity. (e) FMR spectra of the bare YIG (green), YIG/Pt (red) samples. (f) is the YIG thickness

depence of FMR field shift after 3 nm Pt capping along the in-plane (blue) and out-of-plane (red) directions.

**E-field tuning of the magnetic response for the YIG/Pt samples.** Figure 2 displays the FMR response of the YIG/Pt samples before and after applying an electrical bias. The IL gating process was monitored within the ESR cavity, as shown in Figure 2(a). All the FMR measurements were carried out at a microwave frequency of 9.2 GHz. Under a positive $V_g$, the anions (TSFI$^+$) and cations (DEME$^-$) inside IL migrated toward the Pt electrode and Au electrode, respectively. The anions generated an enormous surface charge density up to $10^{15}$ cm$^{-2}$, producing a strong interfacial E-field at the IL/Pt interface. Figure 2(b) demonstrates the $H_r$ shift after applying 4.5 V bias voltage across the IL layer on the GGG/YIG (13 nm)/Pt (3 nm) sample with the external magnetic field parallel to the normal direction of the surface (out of plane). The green line is the initial state and the red line displays the state under 4.5 V $V_g$. In Figure 2(b), the out of plane FMR response shifted 123 Oe toward the high end (larger FMR field). When removing $V_g$, the FMR curve moved towards the intitial state, which is shown in the blue line. The $H_r$ position along with $V_g$ of the same sample is presented in Figure 2(c), and the inset is the schematic of the sample structure. We noticed that only positive E-field improves the MPE and FM ordering. In contrast, negative $V_g$ does not affect the magnetic properties (we did not show here). The gating process was also carried out in the low temperature condition (Figure 4(d)), although the FMR linewidth get broaden at -110 °C, and the $H_r$ shift caused by the gating process increase to 690 Oe, which is 1 order of magnitude higher than the current YIG tunabilities. [13-15]

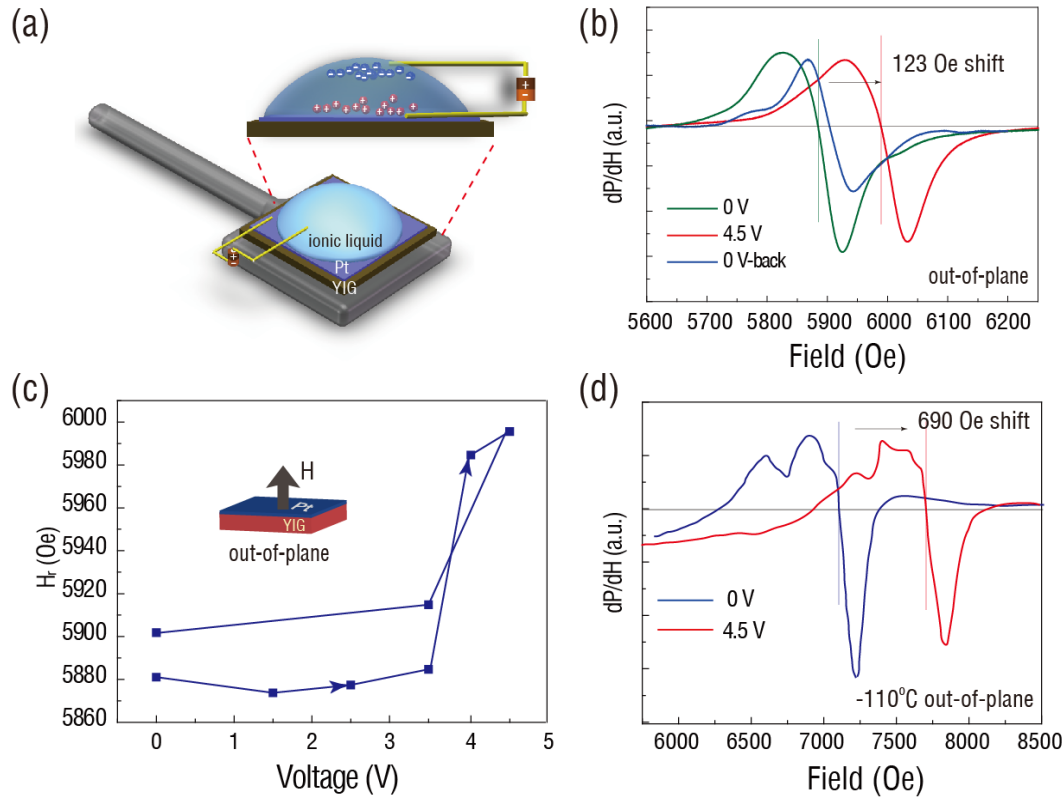

**Figure 2. E-field tuning of the magnetic response for the YIG/Pt samples.** (a) Schematic of the gating process in the FMR cavity. (b) The FMR curves of the YIG (13 nm)/Pt (3 nm) along the out of plane direction under 0 V (green), 4.5 V (red) gating voltage and after remove the gating voltage (blue). (c) Gating voltage dependent of $H_r$ in the YIG (13 nm)/Pt (3 nm) sample along the out of plane direction. (d) The FMR curves of the YIG (13 nm)/Pt (3 nm) along the out of plane direction under 0 V (blue), 4.5 V (red) gating voltage under -115 °C temperature.

We also study the influence of YIG thickness, the external magnetic direction and ambient temperature on the IL gating process. The YIG thickness dependence of the $H_r$ shift under 4.5 V $V_g$ was test. The red line in Figure 3(a) shows the $H_r$ shift in the out of plane direction in different YIG thickness sample. As we can see that similar as the Pt capping result, as represented by the blue curve in Figure 3a, the IL gating caused $H_r$ shift is linear with reciprocal YIG thickness and indicates that IL gating process is essentially an interfacial effect. Figure 3(b) shows the $H_r$ shift of the GGG/YIG (35 nm)/Pt (3 nm) sample under 4.5 V $V_g$ along different external magnetic field direction, where α=0° represents the in-plane direction and α=90° is the out-of-plane direction (α is the angle between the H-field direction and the in-plane direction). In Figure 3(b), the 12 Oe in-plane FMR response shifted toward the low end (smaller FMR field), while the 25 Oe out-of-plane FMR field moved to the high end (larger FMR field). The FMR shifting trend reveals that the in-plane anisotropy at the interface was clearly enhanced, compared with the similar trend of Pt capping (Figure 1(e)). Here, 35 nm YIG is selected for its better RF/microwave signal. Interestingly, as shown in Figure 3(c), a much greater FMR shift of 400 Oe under an out-of-plane magnetic bias was achieved at -110 °C via IL gating (the in-plane FMR shift here is only 22 Oe). The stronger Pt FM ordering and MPE at low temperature may come from smaller thermal perturbations, which also appears in other multiferroic systems such as spin waves in LSMO/PMNPT[33], perpendicular magnetic anisotropy (PMA) structures[34], etc. Besides, the reversibility of FMR switching was also studied in the GGG/YIG (35 nm)/Pt (3 nm) sample along both the out-of-plane directions and in-plane direction, as illustrated in Figure 3d and Figure S4. The $H_r$ was switched back and forth (from 2198 Oe to 2204 Oe in-plane; from 5735 Oe to 5760 Oe out-of-plane) with an alternative E-bias polarity across the IL layer at room temperature.

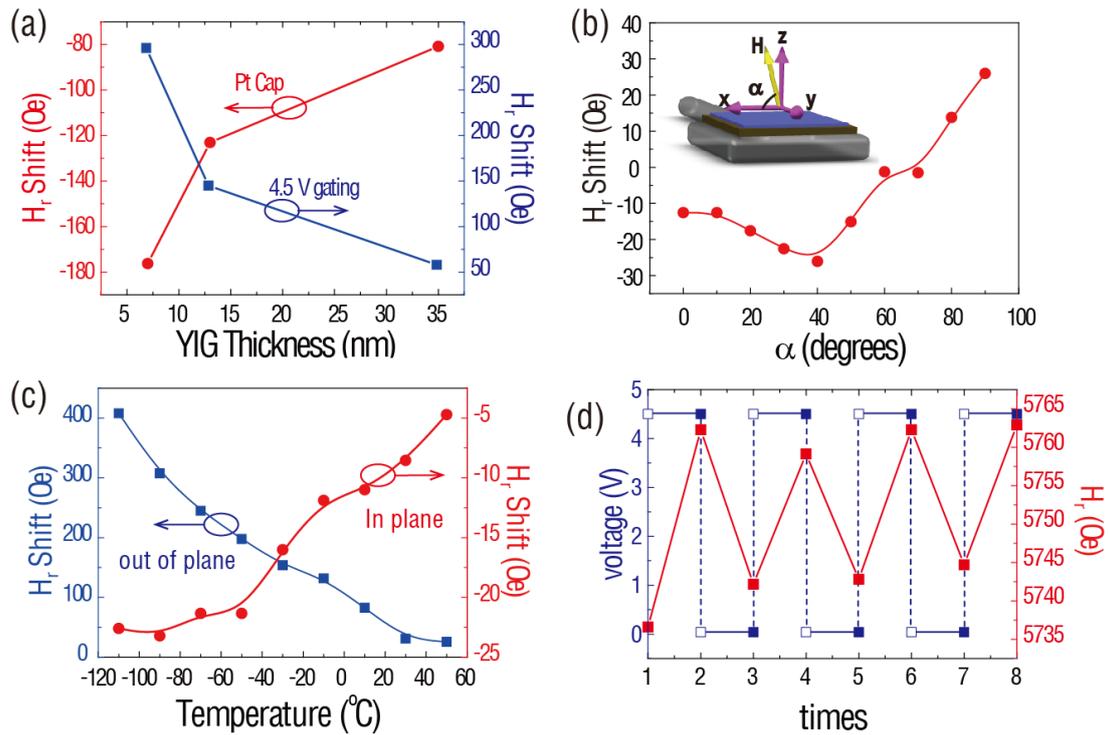

**Figure 3. The influence of the YIG thickness, magnetic field direction, ambient temperature on the tunability of the gating process.** (a) YIG thickness dependence of the FMR shift along the out of plane direction (blue) and in plane direction (red) induced by 4.5 V gating voltage. (b) Angular dependence of FMR field shift induced by 4.5 V $V_g$, inset shows the schematic of the angular dependence between the magnetic field and the film plane. (c) Temperature dependence of the FMR field shift along the out of plane direction (blue) and in plane direction (red) induced by 4.5 V $V_g$. (d) Reproducible test of the gating process in YIG (35 nm)/Pt (3 nm) along the out-of-plane direction at room temperature.

**Discussions**

Many works have proved the existence of the MPE,[7,8,11,35] where several FM ordering atomic layers in normal metals are proximate to an FM material, for example, at the YIG and Pt interface for the YIG/Pt heterostructures. Sun et al. clarified that the FMR curve shift is caused by MPE,[16] and Xiao et al. studied the dependence of the interface structure and the MPE strength by using the first principles calculations based on the density functional theory.[36] They claimed that the FMR shift is originated from the direct exchange interaction between the Fe 3d and Pt 5d electrons via electronic state hybridization and the electron exchange coupling among the Pt atoms. In our experiments, the positive $V_g$ induced cation enrichment at the Pt/IL interface, attracted electrons to the interface and thereby induced an enhanced-MPE-like effect in the YIG/Pt heterostructures. In comparison, Figure S5 shows ionic gating effect in YIG/Cu heterostructure, where the ME tunability can be neglected. We then carry on theoretical calculation to further understand the mechanism under the gating process.

**First Principle calculations on YIG/Pt bilayer system with ionic liquid gating.** To reveal the large enhanced-MPE-like effect, we address first principle calculations to understand what happened after appling the IL gating, as demonstrated in Figure 1(a) with Pt/YIG atomic modeling. In the inset of Figure 4(b), when applying the IL gating, it will generate an E-field along Pt->YIG direction. However, it is hard to build up a chemical potential shift on metals due to the strong screening effect, the electrons of Pt layer will be forced to the Pt/IL interface and leave Pt ion with positive charge on the Pt/YIG interface accordingly. The charge accumulation also builds a back forward E-field to balance the E-field from IL. In this case, numerically, when applying a voltage $V_g$ on IL, the IL will shift the nearest Pt by a energy of $eV_g$, which makes the negative Pt ion ($Pt^{n-}$, $n=V_g$) at the Pt/IL interface. Contrastly, the Pt on the interface between Pt and YIG will be $Pt^{n+}$ for neutralize the total system. Does the ionic Pt has contributions to this enhanced-MPE-like effect? We set up a fcc Pt model, as shown in Figure 4(a), with adding/removing electrons and then calculate the corresponding magnetization of Pt. The spin orbit coupling effect is also considered due to the fact that Pt is a heavy metal. The results are plotted as the blue curve in Figure 4(b). Interestingly, we find that the ionic Pt with around 5 positive charge ($Pt^{5+}$) has suddenly appearred a strong magnetic moment, which is even slightly larger than that of Ni. While applying $V_g$ is around 5 V, we can find a MPE enhancement at the YIG/Pt interface and additional FM ordering in the Pt layer. The estimated gating voltage (5 V) is very close to experimental data (4.5 V), where they show the similar trend of a sudden dump of magnetic enhancement depend on applied voltage, as shown in Figure 4(b). The theoretical ferromagnetic ordering enhancement can be estimated as ~720 Oe, which is very close to the experimental result (690 Oe). In general, the simulation results agree very well with the experiments.

Moreover, we chose $Pt^{5+}$ and $Pt^0$ for detail analysis on the gating effect. As plotted in Figure 4(c) and Figure 4(d), the density of state (DOS) of $Pt^0$ show a strongly energy dependent effect, and the integration of every orbit show that the valance electrons (10 electrons per Pt atom, $5d^9 6s^1$) of $Pt^0$ is mainly d electrons. However the detail numbers are not exactly correct, because we only sum the charge density inside the atomic sphere of Pt while neglect the charge density inside the gap between each atomic spheres. In $Pt^{5+}$ we find that with the Fermi energy being pushed to a much lower energy position, the s and p orbit almost vanished, only d orbit survive, which means that the valance electron of $Pt^{5+}$ is now $5d^5$. we also calculate the spin density for every orbit of $Pt^0$ and $Pt^{5+}$ respectively. The results are plotted in Figure 4(e), Figure 4(f) and Figure S6. It is obvious to see that s and p orbit almost do not give any contribution on magnetization for both $Pt^0$ and $Pt^{5+}$. However, for d orbit, $Pt^0$ and $Pt^{5+}$ show different feathures. $Pt^0$ has a slight spin density which vibrate around zero and end up to a non-magnetic result after summarizing over the total Fermi sea; while $Pt^{5+}$ has a much strong energy dependent effect, which could be three magnitudes larger than that of $Pt^0$ at the some energy level. In the end, the summary over Fermi sea gives a non-trivial magnetization. With Hund rules, it is obviously to have magnetization on every $Pt^{5+}$ atom that can be determined in experiments with strong exchange interacton between $Pt^{5+}$ and magnetic materials. In order to ultilize this interface charge accumulation mechanism, a magnetic insulator (YIG) is a perfect solution in this experiment.

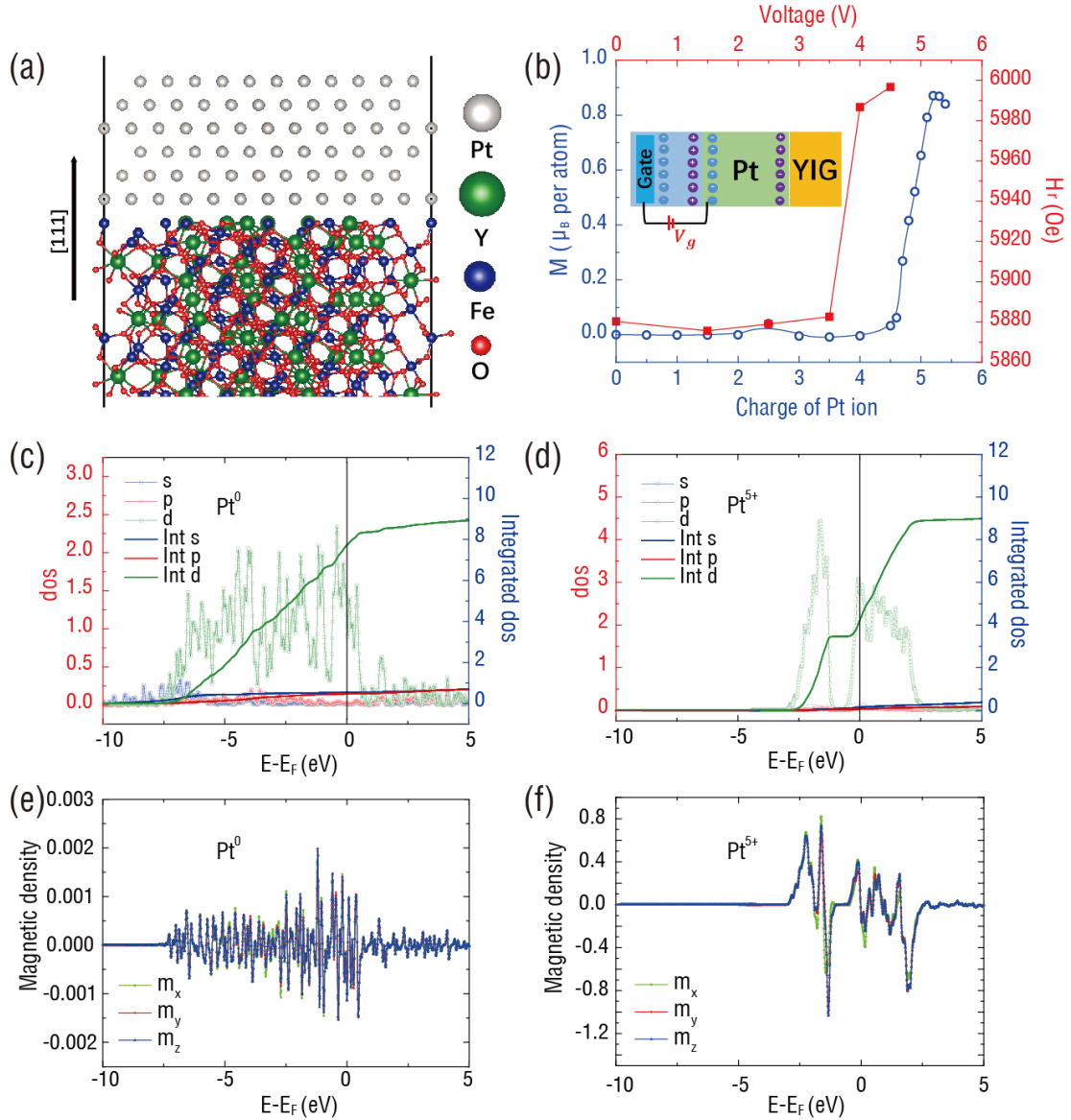

**Figure 4. First Principle calculations on YIG/Pt bilayer system with ionic liquid gating.** (a) The interface model of our calculation. (b) The blue curve is the magnetization of Pt ion as a function of the electron charge of Pt, for example, 5 elctron charge stand for that in the face centre cubic Pt, every atom has been taken 5 electrons away. The red curve is the $H_r$ field along with the increase of the gating voltage. Inset is the schematic of the electron distribution after gating process, here we only show the charge accumulation around the interfaces. (c) (d) reperesnt the density of state (dos) as a function of energy for neutral Pt ($Pt^0$) and 5 electrons positively charged Pt ($Pt^{5+}$)

respectively. And to see more information for the analysis, we seperate them by s- (blue point-line), p- (red point-line), d- (gren point-line) orbit and integrated them over energy to see the occupation for s (blue line), p (red line), d (green line) orbit respectively. (e) (f) The spin density of d orbit for $Pt^0$ and $Pt^{5+}$ respectively. Here $m_x$, $m_y$, $m_z$ stand for the projection of spin density on x, y, z axis.

In summary, based on a novel concept of ionic modulation of magnetic ordering in Pt/YIG bilayer, where the interfacial charge accumulation may enhance the FM ordering of the system and shift the FMR field accordingly, we have realized voltage regulation of YIG thin film by a FET IL gating structure. Outstanding ME tunability up to 690 Oe was achieved in the YIG-based heterostructure, which is 1 orders of magnitude greater than the current YIG tunability, corresponding to a much greater ME FOM of 14. The first principle study revealed a novel E-field induced FM ordering in Pt capping layer and corresponding FMR field tunability via the gating process. This novel IL gating of YIG/heavy metal system is of great research interest and promising for realizing high-performance voltage-tunable YIG based devices.

Method Section

*Sample preparation:* The YIG films for IL gating were deposited on (111) $Gd_3Ga_5O_{12}$ by pulsed laser deposition method. During the deposition, the temperature of substrate was kept at 800 °C while the oxygen pressure was 13 Pa, and the laser pulse rate was 1 Hz. After depositing, the films were annealed in-situ under $5.4\times10^4$ Pa oxygen pressure with the cooling rate of 2 °C/s. After cooling down to room temperature, the YIG films were transferred to the magnetron sputtering chamber. Pt layer was deposited onto these YIG films subsequently.

*Magnetic properties measurements:* Magnetic hysteresis loops of the samples were measured using a LakeShore 7404 vibrating sample magnetometer (VSM). As the magnetization of the YIG films is small (~20 μemu), only in-plane magnetic hysteresis loops of these samples were displayed. Ferromagnetic resonance (FMR) curves of the samples were measured by an X-band electron spin resonance (ESR) system (JOEL, JES-FA200). The magnetic anisotropy change and spin wave patterns were precisely determined.

*Ionic liquid gating preparetion:* We chose the ionic liquid (IL) [DEME]+[TFSI]- as the gating material for its potential tunability and well-studied physicochemical properties. A grid structure, Au/IL/Pt, was formed by using Au and Pt as the gating electrode. Gating voltages from 0 V to 4.5 V were applied to the grid structure using a Keysight B2901A Precision Source/Measure Unit. In the IL phase, the anions and cations migrated toward the Au electrode and the Pt electrode, respectively, driven by the E-field. The charge carrier ions generated an enormous surface charge density up to $10^{15}$ cm$^{-2}$, producing a strong interfacial E-field. The E-field influences on the magnetic properties of these samples were studied by in-situ FMR and VSM measuremants.

During the gating process, low-temperature FMR curves were measured in a cryogenic chamber by liquid $N_2$.

*Structure and morphology analysis:* The crystal structure of the samples as analyzed using a high resolution X-ray diffraction (HRXRD, PANalytical X'Per MRD). The microstructure and morphology of the cross sections of the samples before and after gating process were imaged by high resolution transmission electron microscopy (HRTEM, JEOL JEM-ARM 200F).

Acknowledgements

The work was supported by the Natural Science Foundation of China (Grant Nos. 51472199, 51602244 and11534015), the Natural Science Foundation of Shaanxi Province (Grant No. 2015JM5196), the National 111 Project of China (B14040), and the Fundamental Research Funds for the Central Universities.

The authors appreciate the support from the International Joint Laboratory for Micro/Nano Manufacturing and Measurement Technologies. Z.Z., Z.H and M.L. are supported by the China Recruitment Program of Global Youth Experts. The work at SFU was support by the Natural Science and Engineering Research Council of Canada (NSERC).

Author contributions

M.L., Z.Z. and M.G. conceived and designed the experiments. M.G. and W.S. fabricated samples and carried the in-situ IL-gating control. L.W. carried out the First principle calculations. G.D. did the XRD measurements and the TEM test. All authors contributed to discussion of the results.

**Competing financial interests**

The authors declare no competing financial interests.